\newcommand\Sul{Sul--Cu$_2$Cl$_4$}
\newcommand\IPA{IPA--CuCl$_3$}
\documentclass[prl,twocolumn,floatfix,preprintnumbers,amsmath,amssymb,superscriptaddress]{revtex4}

\usepackage{graphicx}
\usepackage{dcolumn}
\usepackage{bm}

\begin{document}

\title{Quasi-universal finite-$T$ scaling in
gapped one-dimensional quantum magnets.}

\author{A. Zheludev}\email{zheludevai@ornl.gov}
\affiliation{Neutron Scattering Sciences Division, Oak Ridge
National Laboratory, Oak Ridge, Tennessee 37831, USA.}

\author{V.
O. Garlea}
\affiliation{Neutron Scattering Sciences Division, Oak
Ridge National Laboratory, Oak Ridge, Tennessee 37831, USA.}
\author{L.-P.~Regnault}
\affiliation{CEA-Grenoble, DRFMC-SPSMS-MDN, 17 rue des Martyrs,
38054 Grenoble Cedex 9, France.}
\author{H. Manaka}
\affiliation{Graduate School of Science and Engineering, Kagoshima
University, Korimoto, Kagoshima 890-0065, Japan.}
\author{A. Tsvelik}
\affiliation{Brookhaven National Laboratory, Upton, NY 11973,
USA.}
\author{J.-H.~Chung}\altaffiliation{Present address: Department of Physics, Korea University
Anam-dong 5, Seongbook-gu Seoul 126-713, Rep. of Korea.}
 \affiliation{ NCNR, National Institute of Standards and
Technology, Gaithersburg, Maryland 20899, and University of
Maryland, College Park, Maryland, 20742, USA.}

\date{\today}

\begin{abstract}
Temperature dependencies of gap energies and magnon lifetimes are
measured in the quasi-1-dimensional $S=1/2$ gapped quantum magnets
\IPA\ and \Sul\ using inelastic neutron scattering. The results
are compared to those found in literature for $S=1$ Haldane spin
chain materials and to theoretical calculations for the $O(3)$-
and $O(N)$- quantum non-linear $\sigma$-models. It is found that
when the $T=0$ energy gap $\Delta$ is used as the temperature
scale, all experimental and theoretical curves are identical to
within system-dependent but temperature-independent scaling
factors of the order of unity. This quasi-universality extends
over a surprising broad $T$ range, at least up to $\kappa T\sim
1.5\Delta$.
\end{abstract}

\pacs{}

\maketitle

One-dimensional (1D) physics is unique in that collisions between
counter-propagating particles becomes unavoidable, regardless of
their mutual interaction strength. This circumstance has drastic
consequences, such as the collapse of the Landau Fermi Liquid in
1D conductors and  of long range order in gapless spin chains
\cite{Tsvelik}. In these extreme examples the quasiparticle
picture totally breaks down. Not so in {\it gapped} quantum spin
liquids, such as Haldane spin chains \cite{Haldane} or spin
ladders~\cite{Rice1993,Barnes1993}. Here the quasiparticles
(magnons) persist in the limit $T\rightarrow 0$, since mutual
collisions are rare due to their exponentially small density.
Interactions do become important at elevated temperatures thanks
to a thermal excitation of a large number of magnons. The result
is a reduction of magnon lifetimes and a renormalization of their
energies.

In the simplest case the energy gap $\Delta$ is large compared to
the dispersion bandwidth and magnons are localized. The finite-$T$
effects are then well accounted for by the Random Phase
Approximation (RPA) \cite{Leunberger1984,Sasago1997}. This
approach ignores correlations and breaks down if the quasiparticle
are highly mobile, with a velocity $v\gg \Delta$. Fortunately, the
latter regime is open to investigation using field-theoretical
methods. Moreover, as mentioned above, in 1D the details of the
interaction potential become irrelevant. For $\kappa T \ll \Delta$
a universal theory with a single energy scale $\Delta$ can be
derived from the large-$S$ mapping of the Heisenberg Hamiltonian
onto the $O(3)$ non-linear sigma-model (NLSM)
\cite{Sachdev1997,Damle1998}. Until recently, these predictions
and the domain of their applicability remained largely untested
experimentally. Of the strongly 1D gapped quantum magnets,
finite-$T$ data existed only for $S=1$ Haldane spin chains
\cite{Renard88,Zaliznyak1994,Zheludev1996,Kenzelmann2002,Xu2007},
where $\lambda=v/\Delta \sim 6$ \cite{White1993}. In the present
work we explore the temperature dependence of inelastic neutron
spectra in the $S=1/2$ spin ladder system IPA-CuCl$_3$ ($\lambda
\sim 2$) \cite{Masuda2006,Garlea2007,Zheludev2007} and the 4-leg
spin tube Sul-Cu$_2$Cl$_4$ ($\lambda \gtrsim 25$)
\cite{Garlea2008}. We find a striking quasi-universal
renormalization of the magnon gaps and lifetimes that persists to
surprisingly high temperatures.

\begin{figure}
\includegraphics[width=3.5in]{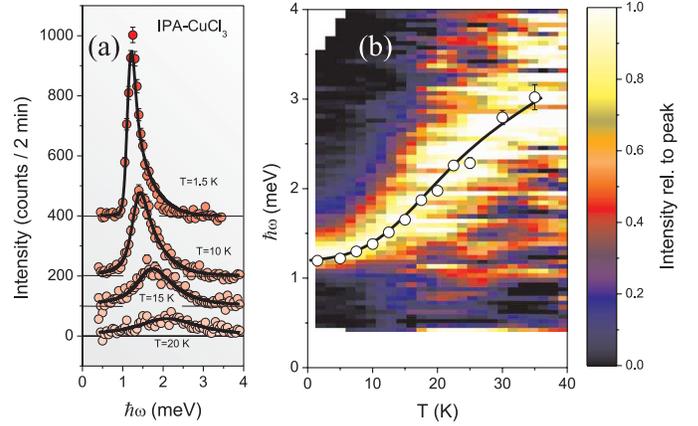}
\caption{ \label{exdata1}(a) Typical background-subtracted
constant-$q$ scans collected at the 1D zone-center $(0.5,0,0)$ in
the spin ladder compound \IPA\ ($v/\Delta=1.9$) at different
temperatures (symbols). Solid lines are fits to the data, as
described in the text. (b) False-color representation of inelastic
intensity measured at the 1D AF zone-center in \IPA\ plotted as a
function of temperature and energy transfer. Symbols are the
measured temperature dependence of the gap energy. The solid line
is a guide for the eye.}
\end{figure}

\begin{figure}
\includegraphics[width=3.5in]{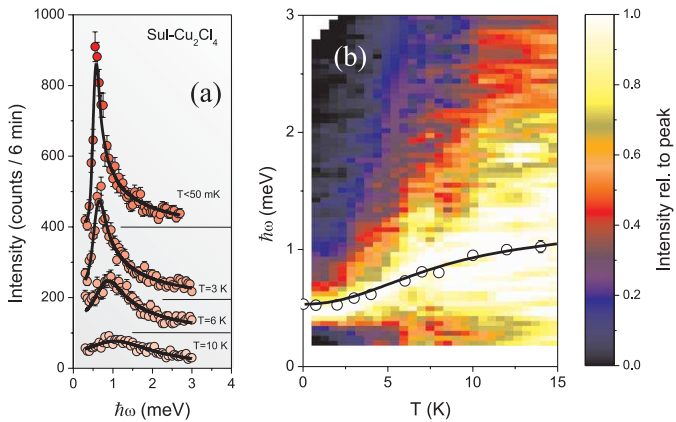}
\caption{ \label{exdata2}Same as Fig.~1, but for the $(0,0,0.5)$
1D 4-leg spin-tube material \Sul\ ($v/\Delta>20$).}
\end{figure}

As described in detail in \cite{Masuda2006}, the magnetic
properties of \IPA\ are due to $S=1/2$ ladders of Cu$^{2+}$ ions
that run along the crystallographic $a$ axis of the  $P
\overline{1}$ crystal structure. Exchange interactions along the
ladder legs are antiferromagnetic (AF). Pairs of spins on each
ladder rung are correlated ferromagnetically. The 1D AF zone
center is at $h=0.5$, where $(h,k,l)$ are components of a wave
vector $\mathbf{q}$. Due to residual interactions between ladders
along the $c$ axis, at low temperatures the spin gap at $h=0.5$
varies between $\Delta=1.17$~meV for $l=0$ to $\Delta=1.8$~meV at
$l=0.5$. There is no detectable dispersion of magnetic excitations
along the $b$ axis. Treating the weak inter-ladder coupling at the
RPA level, one can estimate the intrinsic 1D spin ladder gap
$\Delta_0=1.5$~meV.  The magnon dispersion along the ladder axis
is parabolic at the 1D AF zone-center and  characterized by a
velocity $v=2.9$~meV, as defined by Eq.~2 in \cite{Zheludev2007},
so that $\lambda=v/\Delta_0=1.9$. The other Cu$^{2+}$ based
material that we discuss here is \Sul\ \cite{Garlea2008}. The
corresponding topology of its magnetic interactions is rather
complex, but comes down to that of 4-leg spin tubes with very
strong AF leg coupling along the $c$ axis. Interactions between
adjacent spin tubes are negligible and the system is exceptionally
1D. Due to geometric frustration, the 1D AF zone-center is located
at slightly incommensurate position $l=0.5-\zeta$, where
$\zeta=0.022(2)$.  $\zeta$ amounts to only about 20\% of our
experimental wave vector resolution, and can be safely ignored in
the context of the present study. At low temperatures the gap in
\Sul\  is $\Delta_0=0.55$~meV. Dispersion along the $c^\ast$
direction is very steep, with $v=14$~meV. Thus, for \Sul,
$\lambda=25$ is 12 times larger than for \IPA\ and 4 times larger
than for Haldane chains.

Temperature-dependent inelastic measurements were performed using
3-axis neutron spectrometers. The data on \IPA\ at momentum
transfer $\mathbf{q}=(0.5,0,0)$ were collected on the IN22
instrument at ILL using 5~meV fixed-final neutron energy, a
pyrolitic graphite (PG) monochromator, a horizontally focused PG
analyzer, and a Be filter after the sample. The background was
measured independently at each temperature away from the 1D AF
zone-center, at $\mathbf{q}=(0.3,0,0)$ and $\mathbf{q}=(0.7,0,0)$.
The sample consisted of 20 co-aligned fully deuterated single
crystals of total mass 3~g. Sample environment was a standard
He-flow cryostat. A similar configuration on the NG-5 SPINS
instrument at NIST was used with 3.7~meV fixed final energy
neutrons to collect data around $(0.5,0,0.5)$ in \IPA\ and at
$(0,0,0.5)$ in \Sul. For the latter material, we utilized a $\sim
2$~g assembly of 15 co-aligned deuterated single crystals and a
$^4$He-$^3$He dilution refrigerator.

\begin{figure}
\includegraphics[width=3.2in]{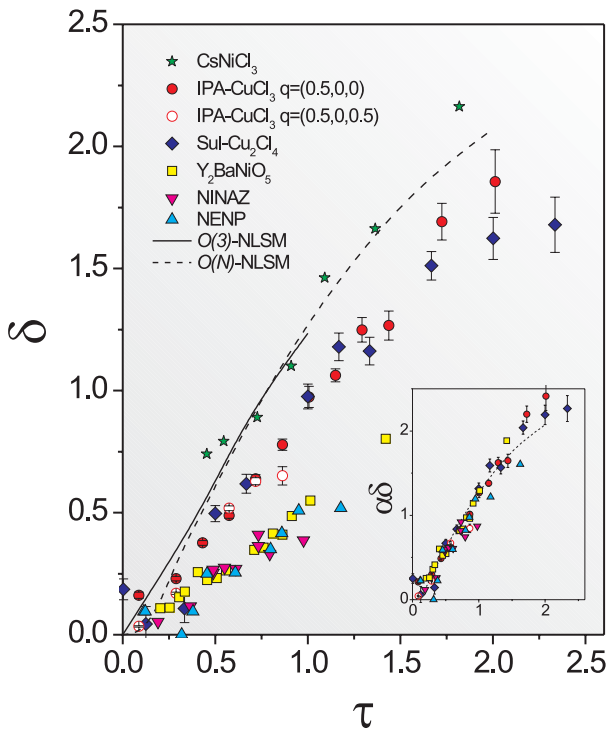}
\caption{ \label{gap} Reduced gap energy $\delta$ plotted against
reduced temperature $\tau$ for diverse quasi 1D quantum
antiferromagnetic materials (symbols) and analytical models
(lines). Inset: same data scaled to show universal behavior.
CsNiCl$_3$, NENP,NINAZ and Y$_2$BasNiO$_5$ data are from
Refs.~\protect\onlinecite{Kenzelmann2002},
\protect\onlinecite{Renard88}, \protect\onlinecite{Zheludev1996}
and \protect\onlinecite{Xu2007}, respectively. Calculations for
the $O(3)$ and $O(N)$ non-linear $\sigma$-models are from
Refs.~\protect\onlinecite{Senechal1993} and
\protect\onlinecite{Jolicoeur1994}, correspondingly.}
\end{figure}

\begin{figure}
\includegraphics[width=3.2in]{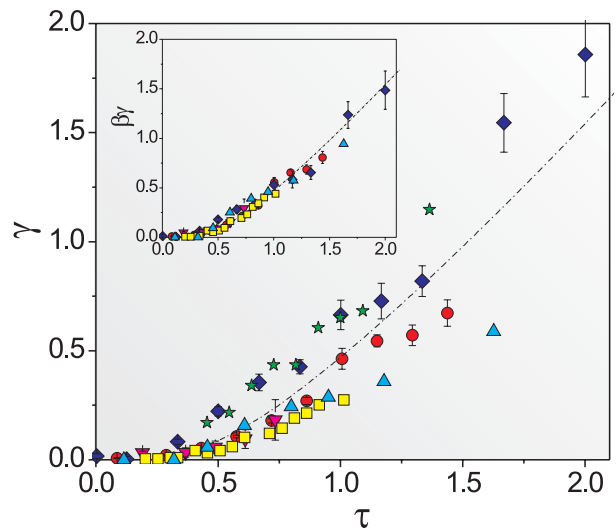}
\caption{ \label{width} Reduced energy width $\gamma$ of the gap
excitations plotted against reduced temperature $\tau$ for diverse
quasi 1D quantum antiferromagnetic materials (symbols, as in
Fig.~\protect\ref{gap}) and the $O(3)$ non-linear $\sigma$-model
(dash-dot line, Ref.~\protect\onlinecite{Damle1998}). }
\end{figure}

Typical background-subtracted energy scans collected at the 1D AF
zone-centers in \IPA\ and \Sul\ at different temperatures are
shown in Fig.~\ref{exdata1}a and Fig.~\ref{exdata2}a,
respectively. Series of such scans are assembled into false-color
plots in Figs~\ref{exdata1}b and Fig.~\ref{exdata2}b. Here the
measured intensity at each temperature was normalized by its peak
value. The observed gap modes broaden with temperature, and the
gap energies increase.  The data were fit to a model cross section
function that was numerically convoluted with the resolution
function of the spectrometer. The dynamic structure factor was
written in the two-Lorentzian form, as given by Eqs.~ 6 and 7 in
\cite{Zaliznyak1994}. Lorentzian line shapes are consistent with
the theoretical results of \cite{Damle1998}. The excitation width
$\Gamma$ was assumed to be $\mathbf{q}$-independent. The
dispersion relation was written as
$[\hbar\omega_0(\mathbf{q})]^2=\Delta^2+v^2(\mathbf{q}\mathbf{d}-\pi)^2$,
with $\mathbf{d}=\mathbf{a}$ for \IPA\ and $\mathbf{d}=\mathbf{c}$
for \Sul, respectively. The parameters $\Gamma$, $\Delta$, and an
overall intensity prefactor were refined to best fit the scans
measured at each temperature. The velocities $v$ were fixed at
their previously determined low-temperature values. Typical fits
are plotted as heavy solid curves in Figs.~\ref{exdata1}a and
~\ref{exdata2}a. The temperature dependencies of the gap energy
$\Delta(T)$ obtained from the fits are plotted in symbols in
Figs.~\ref{exdata1}b and ~\ref{exdata2}b. In Fig.~\ref{gap} we
plot the dimensionless variation of $\Delta$, defined in
quadrature as $\delta(T)=\sqrt{\Delta^2(T)-\Delta^2(0)}/\Delta_0$,
to emphasize the low-temperature region. The abscissa axis shows
the reduced temperature $\tau=\kappa T/\Delta_0$, $\kappa$ being
the Boltzman's constant. The $\tau$-dependence of the relative
excitation half-width $\gamma(T)=\Gamma(T)/\Delta_0$ is shown in
Fig.~\ref{width} \footnote{For $T>25$~K the quality of the data
for \IPA\ was not sufficient to simultaneously fit $\Delta$ and
$\Gamma$ as independent parameters. In that regime only $\Delta$
was refines, while $\Gamma$ was fixed to 1.2~meV.}. Note that for
\IPA\ we obtain a perfect overlap between measurements at the
minimum and maximum of the transverse dispersion, validating our
expectation that the temperature dependence is an intrinsic 1D
effect.

Central to this study is a comparison of our results for \Sul\ and
\IPA\ to those reported in literature for other 1D quantum spin
liquids.  Extensive finite-$T$ data exist for the $S=1/2$ dimer
systems Cu(NO$_3$)$_2$ \cite{Xu2000} and TlCuCl$_3$
\cite{Ruegg2005}. Unfortunately, in the former compound $v\ll
\Delta$, while the latter is by no measure one-dimensional. Among
the known materials that do fit our requirements, neutron data
exist only for $S=1$ Haldane spin chain systems. CsNiCl$_3$ is one
of the oldest known prototypes \cite{Buyers1986,Morra1988}.
Inter-chain interactions in this compound are significant, and it
actually orders in 3D at $T=4.8$~K. At the 3D zone-center the gap
softens at the transition point \cite{Zaliznyak1994}, but one can
hope to retrieve 1D behavior at certain special wave vectors
\cite{Kenzelmann2002}. The intrinsic 1D gap energy in CsNiCl$_3$
is $\Delta(0)\sim 0.9$~meV. In Figs.~\ref{gap} and \ref{width} we
plot the temperature-dependent data from \cite{Kenzelmann2002} in
star symbols. The organo-metallic complexes NENP
\cite{Regnault1994} and NINAZ \cite{Zheludev1996} are excellent 1D
systems, but they are affected by the rather strong magnetic
anisotropy. This anisotropy splits the triplet of Haldane gap
modes into a low energy transverse-polarized doublet and a
longitudinal higher-energy singlet \cite{Golinelli1992}, with gaps
$\Delta_\bot$ and $\Delta_\|$, respectively. As argued in
\cite{Senechal1993}, in this case one can still make a meaningful
comparison with Heisenberg systems by focusing on the two lower
modes. For NENP and NINAZ $\Delta_\bot=1.1$~meV and
$\Delta_\bot=3.6$~meV, respectively. In Figs.~\ref{gap} and
\ref{width} we plot
$\delta(T)=\sqrt{\Delta_\bot^2(T)-\Delta_\bot^2(0)}/\Delta_\bot(0)$
and $\gamma(T)=\Gamma_\bot(T)/\Delta_\bot(0)$ vs.
$\tau=T/\Delta_\bot(0)$, with up and down triangles for NENP
\cite{Renard88} and NINAZ \cite{Zheludev1996}, respectively. The
most recent and extensive study \cite{Xu2007} is on Y$_2$BaNiO$_5$
that is known to be an excellent 1D Haldane gap system with
$\Delta= 8.6$~meV and only weak anisotropy \cite{Xu1996}. In
Figs.~\ref{gap} and \ref{width} the corresponding data from
\cite{Xu2007} are plotted in square symbols. As mentioned above,
for $S=1$ Haldane spin chains numerical calculations predict
$\lambda=6.2$ \cite{White1993}, but it is also useful to quote the
experimentall ratio $\lambda\equiv v/\Delta=6.4$ for CsNiCl$_3$
\cite{Morra1988}, $\lambda=6.1$ for NENP \cite{Regnault1994}, and
$\lambda=8.1$ for Y$_2$BaNiO$_5$ \cite{Xu1996}.

It is instructive to compare the experimental data on \IPA, \Sul\
and the other materials described above to field-theoretical
results. The solid line in Fig.~\ref{gap} shows a one-loop
calculation of the temperature dependence of the gap energy in the
$O(3)$-NLSM from \cite{Senechal1993}. For this model thermal
broadening of excitations was calculated in \cite{Damle1998}, and
is plotted in a dash-dot line in Fig.~\ref{width}. More results
are available for the large-$N$ limit of the NLSM that is more
easily tractable. $O(N)$-NLSM calculations for $\Delta(T)$ were
reported in \cite{Jolicoeur1994} and are plotted in a dashed line
in Fig.~\ref{gap}.

The thermal renormalization of gap energy remains within a factor
of two for all the diverse materials and models considered. From
the theoretical viewpoint, the increase is driven by a repulsion
between thermally excited quasiparticles
\cite{Jolicoeur1994,Sachdev1997,Damle1998}. It is proportional to
the correlation length $\lambda$ and the quasiparticle density
$\rho$. At low temperatures the latter {\it universally} scales as
$\rho \propto \sqrt{\tau}/\lambda\exp{(-1/\tau)}$ for all 1D
systems with a dynamically generated gap. As a result, for $\kappa
T\ll \Delta$ one can expect all measured $\delta(\tau)$
dependencies, as well as the ones calculated for the $O(3)$- and
$O(N)$-NLSM, to be the same, to within a system-dependent but
temperature-independent scaling factor $\alpha$. The latter will
reflect the details of quasiparticle interactions in each
particular model. Experimentally, this is indeed the case in the
entire studied temperature range. The inset in Fig.~\ref{gap}
shows such a scaling plot, where we have arbitrarily selected
$\alpha=1$ for the $O(N)$-NLSM, $\alpha=1.3$ for the spin ladders
in \IPA, $\alpha=1.35$ for \Sul, and $\alpha=2.35$ for all the
Haldane spin chains. The data on CsNiCl$_3$ were excluded from the
scaling, as they are inconsistent with those on other $S=1$
chains, due to the strong 3D interactions in that material
\cite{Zaliznyak1994}.

A similar quasi-universal scaling can be also expected for the
excitation widths. The latter is given by the inverse average time
between collisions among thermally excited quasiparticles
 and is thus proportional to $\rho$ and thermal
average of quasiparticle velocity $\overline{v}\propto
v\sqrt{T/\Delta}$ \cite{Damle1998}. The experimental curves in
Fig.~\ref{width} are indeed very close for all systems under
consideration, and a near-perfect collapse (inset) is obtained
with scaling factors $\beta=1$ for the $O(N)$-NLSM, $\beta=1.2$
for the spin ladders in \IPA, $\beta=0.8$ for \Sul, and
$\beta=1.6$ for Haldane chains.

Deviations from unity of the factors $\alpha$ and $\beta$ reflect
the inadequacy of the NLSM mapping. Indeed, it is rigorous only in
the limit $S>>1$. As evident from recent numerical studies
\cite{White2007}, that model has  only limited quantitative
applicability to $S=1/2$ and $S=1$ chains. An alternative
description of the $S=1/2$ ladder contains quasiparticle
interactions as a free parameter \cite{Shelton1996}, and may be
more versatile. Of course, any quasi-universal behavior with the
energy scale set by $\Delta$ will have a solid theoretical
justification only in the limit $\kappa T\ll\Delta$. A different
universality is expected for $v \gg \kappa T\gg \Delta$, where the
field theoretical description is still justified, but the spin gap
can be considered negligible. Here the only remaining energy scale
is the temperature itself and $\chi"(\pi,\omega)$ is a function of
$\omega/T$.  Most of our data actually lie in the crossover region
$\kappa T\sim \Delta$, while access to the limit $\kappa
T\ll\Delta$ is hindered by the limited experimental resolution.
The two central experimental findings are thus the quasi-universal
$T/\Delta$ scaling in vastly diverse models and real  small-$S$
materials, and its persistence  well beyond the low-temperature
limit, up to at least $\kappa T\sim 1.5 \Delta$.

One of the authors (A.Z.) would like to thank F. Essler, I.
Zanliznyak and Ch. Ruegg for insightful and educational
discussions. Research at ORNL and BNL was funded by the United
States Department of Energy, Office of Basic Energy Sciences-
Materials Science, under Contract No. DE-AC05-00OR22725 and
DE-AC02-98CH10886, respectively. The work at NCNR was supported by
the National Science Foundation under Agreement Nos. DMR-9986442,
-0086210, and -0454672.


\end{document}